\documentclass[aps,prb,preprint]{revtex4}
\usepackage{graphicx,color}
\usepackage{dcolumn}
\usepackage{bm}
\usepackage{amsmath,amsthm,amssymb}

\begin{document}

\title{
Theory of spin transport induced by the temperature gradient
}

\author{Yuu Takezoe, Kazuhiro Hosono, Akihito Takeuchi and Gen Tatara}
\affiliation{%
Department of Physics, Tokyo Metropolitan University,
Hachioji, Tokyo 192-0397, Japan
}
\begin{abstract}
Spin transport driven by the temperature gradient in ferromagnetic metals is studied based on a microscopic theory.
It is shown that the temperature gradient works as an effective field equivalent to the electric field as for both the spin current generation and the spin relaxation torque.
The thermally driven contribution of the spin current and the relaxation torque are thus proportional to $\nabla T$ and $\nabla^2 T$, respectively.
\end{abstract}

\date{\today}

\maketitle
\def\average#1{\left\langle {#1} \right\rangle}
\def\averagefr#1{\left\langle {#1} \right\rangle_0}
\def\avbar#1{\overline{#1}}
\def\impaverage#1{\left\langle {#1} \right\rangle_{\rm i}}
\def\bra#1{\lt\langle{#1} \rt|}
\def\ket#1{\lt|{#1} \rt\rangle}
\def\braket#1#2{\lt.\lt\langle{#1} \rt|{#2}\rt\rangle}
\def\ddelo#1{\frac{d^2}{d #1^2}}
\def\ddel#1#2{\frac{d^2 #1}{d #2 ^2}}
\def\ddelpo#1{\frac{\partial^2}{\partial #1^2}}
\def\delo#1{\frac{d}{d #1}}
\def\delpo#1{\frac{\partial}{\partial #1}}
\def\deldo#1{\frac{\delta}{\delta #1}}
\def\del#1#2{\frac{d #1}{d #2}}
\def\delp#1#2{\frac{\partial #1}{\partial #2}}
\def\deld#1#2{\frac{\delta #1}{\delta #2}}
\def\vvec#1{\stackrel{{\leftrightarrow}}{#1}} 
\def\vectw#1#2{\left(\begin{array}{c} #1 \\ #2 \end{array}\right)}
\def\vecth#1#2#3{\left(\begin{array}{c} #1 \\ #2 \\ #3 \end{array}\right)}
\def\mattw#1#2#3#4{\left(\begin{array}{cc} #1 & #2 \\ #3 & #4 \end{array}\right)}
\def\Eqref#1{Eq. (\ref{#1})}
\def\Eqsref#1#2{Eqs. (\ref{#1})(\ref{#2})}
\def\Eqrefj#1{(\ref{#1})��}
\def\Eqsrefj#1#2{�� (\ref{#1})(\ref{#2})}
\def\dispeq#1{$\displaystyle{#1}$}
\def\ret{{\rm r}}
\def\adv{{\rm a}}
\def\L{{\rm L}}
\def\R{{\rm R}}
\def\green#1#2{g_{#1}^{#2}}
\def\jspin#1#2{j_{{\rm s}#1}^{#2}}
\def\jvspin#1#2{{\jv}_{{\rm s}#1}^{#2}}
\def\ls#1{\ell_{{\rm s}#1}}
\def\listitem#1{\begin{itemize}\item #1 \end{itemize}}
\newcommand{\lt}{\left}
\newcommand{\rt}{\right}
\newcommand{\nablarl}{\stackrel{\leftrightarrow}{\nabla}}
\newcommand{\nnr}{\nonumber\\}
\newcommand{\adag}{{a^{\dagger}}}
\newcommand{\alphaz}{\alpha_0}
\newcommand{\alphasf}{\alpha_\spinflip}
\newcommand{\Area}{A}
\newcommand{\aT}{\overline{\rm T}}
\newcommand{\av}{{\bm a}}
\newcommand{\Av}{{\bm A}}
\newcommand{\Avem}{{\bm A}_{\rm em}}
\newcommand{\Aem}{{A}_{\rm em}}
\newcommand{\Ams}{{\rm A/m}^2}
\newcommand{\Aph}{A^{\phi}}
\newcommand{\Ath}{A^{\theta}}
\newcommand{\Aphv}{\Av^{\phi}}
\newcommand{\Athv}{\Av^{\theta}}
\newcommand{\Az}{{A^{z}}}
\newcommand{\bv}{{\bm b}}
\newcommand{\Bv}{{\bm B}}
\newcommand{\Bz}{{B_z}}
\newcommand{\Bc}{B_{\rm c}}
\newcommand{\Bvs}{{\bm B}_{S}}
\newcommand{\Bveff}{{\bm B}_{\rm eff}}
\newcommand{\Bve}{{\bm B}_{\rm e}}
\newcommand{\betasf}{{\beta_\spinflip}}
\newcommand{\betana}{{\beta_{\rm na}}}
\newcommand{\betaw}{{\beta_{\rm w}}}
\newcommand{\cbar}{\avbar{c}}
\newcommand{\cdag}{{c^{\dagger}}}
\newcommand{\chiz}{\chi^{(0)}}
\newcommand{\chio}{\chi^{(1)}}
\newcommand{\chitilo}{\tilde{\chi}^{(1)}}
\newcommand{\chitilz}{\tilde{\chi}^{(0)}}
\newcommand{\chiuni}{\chi_{0}}
\newcommand{\cH}{c_{H}}
\newcommand{\cHdag}{c_{H}^{\dagger}}
\newcommand{\ckv}{c_{\kv}}
\newcommand{\ckvs}{c_{\kv\sigma}}
\newcommand{\ccv}{{\bm c}}
\newcommand{\Cbeta}{C_\beta}
\newcommand{\Cr}{C_\rightarrow}
\newcommand{\Cl}{C_\leftarrow}
\newcommand{\Ci}{C_{\rm i}}
\newcommand{\Ct}{C_t}
\newcommand{\cv}{{\bm c}}
\newcommand{\ctil}{\tilde{c}}
\newcommand{\dbar}{\avbar{d}}
\newcommand{\deltaS}{\delta S}
\newcommand{\Deltasd}{\Delta_{sd}}
\newcommand{\dels}{{s_0}}
\newcommand{\ddagg}{d^{\dagger}}
\newcommand{\dtil}{\tilde{d}}
\newcommand{\dv}{{\bm d}}
\newcommand{\dw}{{\rm w}}
\newcommand{\dx}{{d^3 x}}
\newcommand{\Deltatil}{\tilde{\Delta}}
\newcommand{\Dcal}{{\cal D}}
\newcommand{\DOS}{N}
\newcommand{\dos}{\nu}
\newcommand{\doss}{\nu_0}
\newcommand{\dosu}{\dos_{+}}
\newcommand{\dosd}{\dos_{-}}
\newcommand{\dosp}{\dos_{+}}
\newcommand{\dosm}{\dos_{-}}
\newcommand{\DOSV}{{N(0)}}
\newcommand{\DOSom}{{\DOS_\omega}}
\newcommand{\ef}{{\epsilon_F}}
\newcommand{\eF}{{\epsilon_F}}
\newcommand{\eft}{{\epsilon_F \tau}}
\newcommand{\eftoh}{\frac{\epsilon_F \tau}{\hbar}}
\newcommand{\eftauinv}{\frac{\hbar}{\epsilon_F \tau}}
\newcommand{\ekv}{\epsilon_{\kv}}
\newcommand{\ekvs}{\epsilon_{\kv\sigma}}
\newcommand{\ekvp}{\epsilon_{\kv'}}
\newcommand{\ekvps}{\epsilon_{\kv'\sigma}}
\newcommand{\elld}{{\ell_{\rm D}}}
\newcommand{\ells}{\ell_{\rm s}}
\newcommand{\Ev}{{\bm E}}
\newcommand{\ev}{{\bm e}}
\newcommand{\evth}{{\bm e}_{\theta}}
\newcommand{\evph}{{\bm e}_{\phi}}
\newcommand{\evs}{{\bm n}}
\newcommand{\evsz}{{\evs}_0}
\newcommand{\evsph}{(\evph\times\evz)}
\newcommand{\evx}{{\bm e}_{x}}
\newcommand{\evy}{{\bm e}_{y}}
\newcommand{\evz}{{\bm e}_{z}}
\newcommand{\fl}{{\eta}}
\newcommand{\fltil}{\tilde{\eta}}
\newcommand{\flitil}{\tilde{\fl_{\rm I}}}
\newcommand{\flrtil}{\tilde{\fl_{\rm R}}}
\newcommand{\fli}{{\fl_{\rm I}}}
\newcommand{\flr}{{\fl_{\rm R}}}
\newcommand{\fkv}{f_{\kv}}
\newcommand{\fkvs}{f_{\kv\sigma}}
\newcommand{\fo}{{f(\omega)}}
\newcommand{\fpo}{{f'(\omega)}}
\newcommand{\fomega}{{\omega}}
\newcommand{\fbeta}{f^{\beta}}
\newcommand{\fpin}{{f_{\rm pin}}}
\newcommand{\Fpin}{{F_{\rm pin}}}
\newcommand{\fe}{{f_{\rm e}}}
\newcommand{\fna}{{f_{\rm ref}}}
\newcommand{\Fe}{F}
\newcommand{\Fev}{\Fv}
\newcommand{\Fbeta}{F^{\beta}}
\newcommand{\Fbetav}{\Fv^{\beta}}
\newcommand{\Fbetafactor}{\mu}
\newcommand{\Fhall}{F^{\rm Hall}}
\newcommand{\Fhallv}{\Fv^{\rm Hall}}
\newcommand{\Fren}{F^{\rm ren}}
\newcommand{\Frenv}{\Fv^{\rm ren}}
\newcommand{\Fvna}{\Fv^{\rm ref}}
\newcommand{\Fnav}{\Fv^{\rm ref}}
\newcommand{\Fna}{F^{\rm ref}}
\newcommand{\Fad}{\Fhall}
\newcommand{\Fzad}{F^{\rm (0)ad}}
\newcommand{\Fv}{{\bm F}}
\newcommand{\Fo}{F^{(1)}}
\newcommand{\Fw}{F_{\rm w}}
\newcommand{\Fz}{F^{(0)}}
\newcommand{\Fzv}{\Fv^{(0)}}
\newcommand{\Fdelta}{\delta F}
\newcommand{\Fdeltav}{\delta \Fv}
\newcommand{\Fdel}{\delta \Fo}
\newcommand{\Fdelv}{\delta \Fv^{(1)}}
\newcommand{\Gv}{{\bm G}}
\newcommand{\gv}{{\bm g}}
\newcommand{\gr}{g^{\rm r}}
\newcommand{\gto}{g^{\rm t}}
\newcommand{\ga}{g^{\rm a}}
\newcommand{\Gr}{G^{\rm r}}
\newcommand{\Ga}{G^{\rm a}}
\newcommand{\Gto}{G^{\rm t}}
\newcommand{\Gat}{G^{\overline{\rm t}}}
\newcommand{\Gless}{G^{<}}
\newcommand{\Ggrt}{G^{>}}
\newcommand{\gless}{g^{<}}
\newcommand{\ggrt}{g^{>}}
\newcommand{\Gtil}{\tilde{G}}
\newcommand{\Gcal}{{\cal G}}
\newcommand{\gap}{\Delta_{\rm sw}}
\newcommand{\gammap}{\gamma_{+}}
\newcommand{\gammam}{\gamma_{-}}
\newcommand{\gammaz}{\gamma_{0}}
\newcommand{\grst}{|0\rangle}
\newcommand{\grsthc}{\langle0|}
\newcommand{\grft}{|\ \rangle}
\newcommand{\gyro}{\gamma}
\newcommand{\gyroz}{\gyro_{0}}
\newcommand{\hf}{\frac{1}{2}}
\newcommand{\HA}{{H_{A}}}
\newcommand{\HB}{H_{B}}
\newcommand{\Hv}{\bm{H}}
\newcommand{\He}{H_{\rm e}}
\newcommand{\Heff}{H_{\rm eff}}
\newcommand{\Hem}{H_{\rm em}}
\newcommand{\Hex}{H_{\rm ex}}
\newcommand{\Himp}{H_{\rm imp}}
\newcommand{\Hint}{H_{\rm int}}
\newcommand{\HR}{{H_{\rm R}}}
\newcommand{\Hs}{{H_{\rm S}}}
\newcommand{\Hsf}{{H_\spinflip}}
\newcommand{\Hso}{{H_{\rm so}}}
\newcommand{\Hsd}{H_{sd}}
\newcommand{\Hst}{{H_{\rm ST}}}
\newcommand{\Hw}{H_{\dw}}
\newcommand{\Hz}{H_{0}}
\newcommand{\hbarinv}{\frac{1}{\hbar}}
\renewcommand{\Im}{{\rm Im}}
\newcommand{\ime}{\gamma}
\newcommand{\intinf}{\int_{-\infty}^{\infty}}
\newcommand{\intek}{\int_{-\ef}^{\infty}\! d\epsilon}
\newcommand{\intom}{\int\! \frac{d\omega}{2\pi}}
\newcommand{\intx}{\int\! {d^3x}}
\newcommand{\intk}{\int\! \frac{d^3k}{(2\pi)^3}}
\newcommand{\intr}{\int\! {d^3r}}
\newcommand{\intt}{\int_{-\infty}^{\infty}\! {dt}}
\newcommand{\ioh}{\frac{i}{\hbar}}
\newcommand{\iv}{\bm{i}}
\newcommand{\Ibar}{\overline{I}}
\newcommand{\Iv}{\bm{I}}
\newcommand{\Jex}{{J_{\rm ex}}}
\newcommand{\Jsd}{J_{sd}}
\newcommand{\Js}{{J_{\rm s}}}
\newcommand{\Jv}{\bm{J}}
\newcommand{\js}{j_{\rm s}}
\newcommand{\jsc}{{j_{\rm s}^{\rm c}}}
\newcommand{\jsv}{\bm{j}_{\rm s}}
\newcommand{\Jsv}{\bm{J}_{\rm S}}
\newcommand{\JSv}{\bm{J}_{\rm S}}
\newcommand{\JS}{J_{\rm S}}
\newcommand{\JStotv}{\bm{J}_{S,{\rm tot}}}
\newcommand{\jc}{j_{\rm c}}
\newcommand{\jci}{{{j}_{\rm c}^{\rm i}}}
\newcommand{\jce}{{{j}_{\rm c}^{\rm e}}}
\newcommand{\jatil}{{\tilde{j}_{\rm a}}}
\newcommand{\jctil}{{\tilde{j}_{\rm c}}}
\newcommand{\jcitil}{{\tilde{j}_{\rm c}^{\rm i}}}
\newcommand{\jcetil}{{\tilde{j}_{\rm c}^{\rm e}}}
\newcommand{\jstil}{{\tilde{j}_{\rm s}}}
\newcommand{\jtil}{{\tilde{j}}}
\newcommand{\jv}{\bm{j}}
\newcommand{\kB}{{k_B}}
\newcommand{\kb}{{k_B}}
\newcommand{\kv}{{\bm k}}
\newcommand{\kvxv}{\kv\cdot\xv}
\newcommand{\kvo}{{\kv_1}}
\newcommand{\kvp}{{\kv}'}
\newcommand{\kpq}{{k+\frac{q}{2}}}
\newcommand{\kmq}{{k-\frac{q}{2}}}
\newcommand{\kvpq}{{\kv}+\frac{\qv}{2}}
\newcommand{\kvmq}{{\kv}-\frac{\qv}{2}}
\newcommand{\kvopq}{{\kvo}+\frac{\qv}{2}}
\newcommand{\kvomq}{{\kvo}-\frac{\qv}{2}}
\newcommand{\kvppq}{{\kvp}+\frac{\qv}{2}}
\newcommand{\kvpmq}{{\kvp}-\frac{\qv}{2}}
\newcommand{\kf}{{k_F}}
\newcommand{\kF}{{k_F}}
\newcommand{\kfpm}{{k_{F\pm}}}
\newcommand{\kfmp}{{k_{F\mp}}}
\newcommand{\kfp}{k_{F+}}
\newcommand{\kfm}{k_{F-}}
\newcommand{\kfu}{k_{F+}}
\newcommand{\kfd}{k_{F-}}
\newcommand{\kfs}{k_{F\sigma}}
\newcommand{\kom}{{k_\omega}}
\newcommand{\Kp}{{K_\perp}}
\newcommand{\ktil}{\tilde{k}}
\newcommand{\lams}{{\lambda_{\rm s}}}
\newcommand{\lamv}{{\lambda_{\rm v}}}
\newcommand{\lamso}{{\lambda_{\rm so}}}
\newcommand{\lamz}{{\lambda_{0}}}
\newcommand{\Le}{{L_{\rm e}}}
\newcommand{\Lez}{{L_{\rm e}^0}}
\newcommand{\Leff}{L_{\rm eff}}
\newcommand{\Lb}{L_{\rm B}}
\newcommand{\Ldw}{L_{\dw}}
\newcommand{\Lsd}{{L_{sd}}}
\newcommand{\Ls}{{L_{\rm S}}}
\newcommand{\Lsw}{L_{\rm sw}}
\newcommand{\Lswdw}{L_{\rm sw-dw}}
\newcommand{\Linv}{{\frac{1}{L}}}
\newcommand{\lstil}{\tilde{l_\sigma}}
\newcommand{\Lv}{{\bm L}}
\newcommand{\mv}{{\bm m}}
\newcommand{\Mv}{{\bm M}}
\newcommand{\Mphi}{{M_{\phi}}}
\newcommand{\Mw}{{M_{\dw}}}
\newcommand{\Ms}{M_{\rm s}}
\newcommand{\MR}{{\rm MR}}
\newcommand{\Mz}{M_{0}}
\newcommand{\mus}{{g\mu_{B}}}
\newcommand{\mub}{\mu_B}
\newcommand{\muB}{\mu_B}
\newcommand{\muz}{\mu_0}
\newcommand{\muu}{\mu_+}
\newcommand{\mud}{\mu_-}
\newcommand{\muspin}{\mu_{\rm s}}
\newcommand{\nel}{n_{\rm e}}
\newcommand{\Ne}{N_{\rm e}}
\newcommand{\nv}{{\bm n}}
\newcommand{\Nimp}{N_{\rm imp}}
\newcommand{\nimp}{n_{\rm imp}}
\newcommand{\Ninv}{\frac{1}{N}}
\newcommand{\Nw}{N_{\dw}}
\newcommand{\nvortex}{n_{\rm v}}
\newcommand{\nvz}{{\nv}_0}
\newcommand{\nso}{n_{\so}}
\newcommand{\np}{{n}_+}
\newcommand{\nm}{{n}_-}
\newcommand{\nz}{{n}_0}
\newcommand{\om}{{\omega}}
\newcommand{\omegap}{\omega'}
\newcommand{\Omegatil}{\tilde{\Omega}}
\newcommand{\Omegap}{\Omega'}
\newcommand{\Omegapin}{\Omega_{\rm pin}}
\newcommand{\ompOm}{{\omega+\frac{\Omega}{2}}}
\newcommand{\ommOm}{{\omega-\frac{\Omega}{2}}}
\newcommand{\Omz}{\Omega_0}
\newcommand{\ompOmz}{{\omega+\frac{\Omz}{2}}}
\newcommand{\ommOmz}{{\omega-\frac{\Omz}{2}}}
\newcommand{\Omhf}{\frac{\Omega}{2}}
\newcommand{\phiz}{{\phi_0}}
\newcommand{\Phiv}{\bm{\Phi}}
\newcommand{\PhiB}{{\Phi_{\rm B}}}
\newcommand{\Ptil}{{\tilde{P}}}
\newcommand{\pv}{{\bm p}}
\newcommand{\Pv}{{\bm P}}
\newcommand{\PDOS}{P_{\DOS}}
\newcommand{\qv}{{\bm q}}
\newcommand{\qvxv}{\qv\cdot\xv}
\newcommand{\qtil}{{\tilde{q}}}
\newcommand{\ra}{\rightarrow}
\renewcommand{\Re}{{\rm Re}}
\newcommand{\rhow}{{\rho_{\dw}}}
\newcommand{\rhos}{{\rho_{\rm s}}}
\newcommand{\rhoS}{{\rho_{\rm s}}}
\newcommand{\rhoxy}{{\rho_{xy}}}
\newcommand{\RS}{{R_{\rm S}}}
\newcommand{\Rw}{{R_{\dw}}}
\newcommand{\rv}{{\bm r}}
\newcommand{\Rv}{{\bm R}}
\newcommand{\sd}{$s$-$d$}
\newcommand{\sigmav}{{\bm \sigma}}
\newcommand{\sigmaB}{\sigma_{\rm B}}
\newcommand{\sigmab}{\sigma_{\rm B}}
\newcommand{\sigmaz}{\sigma_0}
\newcommand{\sigmas}{\sigma_{\rm s}}
\newcommand{\se}{{s}}
\newcommand{\sev}{{\bm \se}}
\newcommand{\sevsf}{{\bm \se}_\spinflip}
\newcommand{\SE}{\Sigma}
\newcommand{\SEr}{\Sigma^{\rm r}}
\newcommand{\SEa}{\Sigma^{\rm a}}
\newcommand{\SEless}{\Sigma^{<}}
\newcommand{\sgn}{{\rm sgn}}
\newcommand{\sz}{{s}_0}
\newcommand{\sv}{{{\bm s}}}
\newcommand{\seth}{{\se}_\theta}
\newcommand{\seph}{{\se}_\phi}
\newcommand{\sez}{{\se}_z}
\newcommand{\so}{{\rm so}}
\newcommand{\spol}{{M}}
\newcommand{\spinflip}{{\rm sr}}
\newcommand{\svtil}{\tilde{\bm s}}
\newcommand{\stil}{\tilde{\se}}
\newcommand{\stilz}{\stil_{z}}
\newcommand{\stilpm}{\stil^{\pm}}
\newcommand{\stilpmz}{\stil^{\pm(0)}}
\newcommand{\stilpma}{\stil^{\pm(1{\rm a})}}
\newcommand{\stilpmb}{\stil^{\pm(1{\rm b})}}
\newcommand{\stilpmo}{\stil^{\pm(1)}}
\newcommand{\stilpara}{\stil_{\parallel}}
\newcommand{\stilperp}{\stil_{\perp}}
\newcommand{\Simpv}{{{\bm S}_{\rm imp}}}
\newcommand{\Simp}{{S_{\rm imp}}}
\newcommand{\Stot}{{S_{\rm tot}}}
\newcommand{\Stotv}{\bm{S}_{\rm tot}}
\newcommand{\Sh}{{\hat {S}}}
\newcommand{\Svh}{{\hat {\Sv}}}
\newcommand{\Sv}{{{\bm S}}}
\newcommand{\Svz}{{{\bm S}_0}}
\newcommand{\sumx}{{\int\! \frac{d^3x}{a^3}}}
\newcommand{\sumk}{{\sum_{k}}}
\newcommand{\sumkv}{{\sum_{\kv}}}
\newcommand{\sumqv}{{\sum_{\qv}}}
\newcommand{\sumom}{\int\!\frac{d\omega}{2\pi}}
\newcommand{\sumOm}{\int\!\frac{d\Omega}{2\pi}}
\newcommand{\sumomOm}{\int\!\frac{d\omega}{2\pi}\int\!\frac{d\Omega}{2\pi}}
\newcommand{\thickness}{{d}}
\newcommand{\thetaz}{{\theta_0}}
\newcommand{\tr}{{\rm tr}}
\newcommand{\To}{{\rm T}}
\newcommand{\Ta}{\overline{\rm T}}
\newcommand{\Tc}{{\rm T}_{C}}
\newcommand{\Tct}{{\rm T}_{\Ct}}
\newcommand{\tcmp}{\tau}
\newcommand{\tcmpi}{\tau_{\rm I}}
\newcommand{\tcmpinf}{\tau_{\infty}}
\newcommand{\tcmpz}{\tau_{0}}
\newcommand{\tcmpzp}{\tau_{0}'}
\newcommand{\Torqv}{{\bm \tau}}
\newcommand{\torque}{{\tau}}
\newcommand{\torquev}{{\bm \torque}}
\newcommand{\Torquev}{{\bm \tau}}
\newcommand{\torqueve}{\torquev}
\newcommand{\torquee}{\torque}
\newcommand{\torquew}{{\torque_{\dw}}}
\newcommand{\tautil}{{\tilde{\tau}}}
\newcommand{\taup}{\tau_{+}}
\newcommand{\taum}{\tau_{-}}
\newcommand{\tauw}{\tau_{\dw}}
\newcommand{\taus}{\tau_{\rm s}}
\newcommand{\tausf}{\tau_\spinflip}
\newcommand{\thetast}{\theta_{\rm st}}
\newcommand{\ttil}{{\tilde{t}}}
\newcommand{\tinf}{t_\infty}
\newcommand{\tz}{t_0}
\newcommand{\Ubar}{\overline{U}}
\newcommand{\Ueff}{U_{\rm eff}}
\newcommand{\Uz}{U_0}
\newcommand{\Uv}{U_V}
\newcommand{\vc}{{v_{\rm c}}}
\newcommand{\ve}{{v_{\rm e}}}
\newcommand{\vev}{{\vv_{\rm e}}}
\newcommand{\vv}{\bm{v}}
\newcommand{\vs}{{v_{\rm s}}}
\newcommand{\vsv}{{\vv_{\rm s}}}
\newcommand{\vw}{v_{\rm w}}
\newcommand{\vf}{{v_F}}
\newcommand{\vimp}{v_{\rm imp}}
\newcommand{\vi}{{v_{\rm i}}}
\newcommand{\Vi}{{V_{\rm i}}}
\newcommand{\Vso}{v_{\rm so}}
\newcommand{\vtil}{{{v_0}}}
\newcommand{\Vpin}{{V}_{\rm pin}}
\newcommand{\Vinv}{\frac{1}{V}}
\newcommand{\vz}{{v_0}}
\newcommand{\Vz}{{V_0}}
\newcommand{\Vcal}{{\cal V}}
\newcommand{\Vztil}{{\tilde{V_0}}}
\newcommand{\Ws}{{W_{\rm S}}}
\newcommand{\Xtil}{{\tilde{X}}}
\newcommand{\xv}{{\bm x}}
\newcommand{\Xv}{{\bm X}}
\newcommand{\xvp}{{\bm x}_{\perp}}
\newcommand{\xw}{{z}}
\newcommand{\Xz}{{X_0}}
\newcommand{\Zs}{Z_{\rm S}}
\newcommand{\Zz}{Z_{0}}
\newcommand{\ztil}{u}
\newcommand{\zh}{\hat{z}}

\def\ret{{\rm r}}
\def\adv{{\rm a}}
\def\eks{\epsilon_{k\sigma}}
\def\ek{\epsilon_{k}}
\def\ekp{\epsilon_{k'}}
\def\ekps{\epsilon_{k'\sigma}}
\def\kvperp{{\bm k}_{\perp}}
\def\L{{n}}
\def\R{{m}}
\def\green#1#2{g_{#1}^{#2}}

\section{Introduction}

Thermoelectric effects such as the Seebeck and the Nernst effects
have been studied for more than a hundred years, and are applied to various devices like thermocouples.
The effects have been successfully explained by phenomenological and microscopic theories \cite{Ashcroft76,Berger72}, as for the charge transports.

Recently, spintronics, which aims at the control of the electron spins, is attracting special attention from the viewpoints of the fundamental science and application. 
Of particular importance in the spintronics is the spin current.
The spin current is generated by applying the electric field \cite{Hirsch99}
 or by use of the magnetization dynamics via the spin pumping effect \cite{Silsbee79,Tserkovnyak02}.
Detection of the spin current can be carried out electrically by use of the inverse spin Hall effect \cite{Saitoh06}, which converts the spin current into the charge current or electric voltage using the spn-orbit interaction.
In 2008, another method to creat the spin current was discovered by Uchida et al., namely, the spin Seebeck effect \cite{Uchida08}.
They have applied a temperature gradient to a ferromagnetic metal (permalloy) under the condition that no charge current flows.
By attaching a Pt contact on the permalloy in the perpendicular direction, they found that there appears a finite voltage across the Pt lead.
Since no charge current flows in the permalloy, Uchida et al. concluded that the voltage is the result of the inverse spin Hall effect due to the spin current that is induced by the temperature gradient in the permalloy.
Uchida et al. thus demonstrated that the spin current can be induced by the temperature gradient similarly to the Seebeck effect for the charge.

Thermal effects on magnetic domain walls such as the eddy current induced at the domain wall due to the Nernst-Ettingshausen effect were discussed by Berger \cite{Berger79,Berger85,Jen_a86}.
Very recently, theoretical studies on the thermally-driven spintronics phenomena have been intensively carried out \cite{Hatami07,Kovalev09,Bauer10}.
Thermal spin-transfer torque was discussed by Hatami et al \cite{Hatami07}, and its inverse effect was argued by Kovalev et al. \cite{Kovalev09}.
An unified description of magnetic, electric, thermal and mechanical forces was presented by Bauer et al. \cite{Bauer10}.

In the conventional (charge) Seebeck effect, the effective electric field $E$ is induced proportional to the temperature gradient, $\nabla T$,  as
$E=S\nabla T$, where $S$ is called the Seebeck coefficient. 
As general argument indicates that the Seebeck coefficient of free electrons at low temperatures is written by a energy derivative of the electric conductivity $\sigma_{\rm B}(\epsilon)$ as  \cite{Ashcroft76}
\begin{align}
S= \frac{\pi^2}{3e}(k_{\rm B})^2 T
\lt. \frac{ \frac{d\sigma_{\rm B}(\epsilon)}{d\epsilon} } {\sigma_{\rm B}(\epsilon)}  \rt|_{\epsilon=\mu}, \label{Seebeck0}
\end{align}
 where $\mu$ is the chemical potential.�
In ferromagnets, the exisitence of the charge current indicates also that of the spin current, since the conduction electrons are spin polarized.
Defining the spin current in a uniform ferromagnet as 
$\js\equiv j_{+}-j_-$, where $j_{\pm}$ denotes the current carried by the electron with spin $\pm$.
Treating the two spin channels as independent, thermally induced spin current reads from \Eqref{Seebeck0}
\begin{align}
\js=
 \frac{\pi^2}{3e}(k_{\rm B})^2 T \sum_{\pm}(\pm)
\lt. \frac{d\sigma_{{\rm B,\pm}}(\epsilon)} {d\epsilon} \rt|_{\epsilon=\mu} \nabla T, \label{spinSeebeck0}
\end{align}
where $\sigma_{{\rm B},\pm}$ represents the conductivit for the spin $\pm$ electron.
We define   
the spin Seebeck coefficient $S_s$ as 
$\js=\sigma_{s} S_s \nabla T$, where $\sigma_s\equiv \sigma_{{\rm B},+}-\sigma_{{\rm B},-}$ is the spin cunductivity.
It then reads
\begin{align}
S_s = \frac{\pi^2}{3e}(k_{\rm B})^2 T
\lt. \frac{ \sum_{\pm} (\pm) \frac{d\sigma_{{\rm B},\pm}(\epsilon)}{d\epsilon} } 
 {\sum_{\pm} (\pm) \sigma_{{\rm B},\pm}(\epsilon)} 
 \rt|_{\epsilon=\mu}. \label{spinSeebeck}
\end{align}
The aim of this paper is to derive this expression on a microscoic model, and to extend the argument to a general case with inhomogeneous magnetization, and to study the spin relaxation torque.

Most crucial feature of the spin current when compared to the charge current is the violation of conservation law in solids. 
The spin density $\sv$ and the spin current density $\jsv$  thereby satisfy the continuity equation
\begin{align}
\dot{s}^\alpha+\nabla \cdot\jsv^\alpha={\cal T}^\alpha,  \label{sconteq}
\end{align}
where $\alpha=x,y,z$ is the spin index and ${\cal T}$ is the spin relaxation torque resulting in the non-conservation of the spin.
In metals, the dominant origin of ${\cal T}$ is the spin-orbit interaction.
The relaxation torque has been treated by introducing a phenomenological spin chemical potential and the spin relaxation time \cite{Son87,Valet93}.
The relaxation torque plays essential roles in spintronics phenomena such as the current-induced magnetization switching \cite{TKS_PR08} and the invserse spin Hall effects \cite{Takeuchi10}.
Thus microscopic study of the torque is urgent and important.
The spin relaxation torque induced electrically was recently studied microscopically \cite{Nakabayashi10}, but the thermal contribution has not been explored.

The aim of this paper is to theoretically study the spin transport induced by the temperature gradient.
The temperature gradient is modeled by considering a system made up of subsystems having different temperatures and chemical potentials. Each subsystem is assumed to be in local equilibrium.
The electron transport is studied by introducing the electron hopping between subsystems.
For spin current, we also take into account the inhomogenuity of the magnetization, up to the first order in the spatial derivative. 
The spin relaxation torque is studied in the homogeneous magnetization case and in the presence of the spin-orbit interaction due to random impurities.
In the context of charge transport, the temperature gradient has been known to act as an effective electric field \cite{Ashcroft76}.
We will show the equivalence of $\nabla T$ and the electric field holds also in the spin transports and in the spin relaxation phenomena.

Although the spin-orbit interaction is essential in studying the spin relaxation torque, we will neglect it in studying the spin current, since we are interested in how the temperature gradient acts as an driving force on the dominant spin current but not in deriving the full transport equation.
In fact, the spin-orbit correction to the spin current has been known to have the same dependence on the driving field as the contribution without the spin-orbit interaction \cite{Nakabayashi10}.

\section{Model}

\begin{figure}[tbh]
\begin{center}
\includegraphics[width=0.5\hsize]{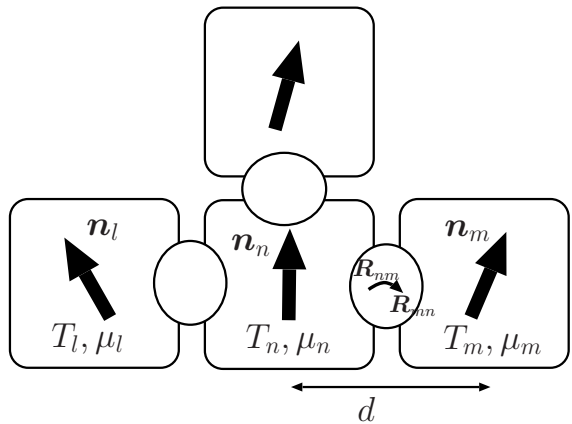}
\caption{ 
The discrete model we consider, made up of subsystem labeled by $l,m,n\cdots$ connected by leads.
Each subsystem $n$ is assumed to be in the local equilibrium at temperature  $T_n$ and the chemical potential $\mu_n$, and have a uniform magentization $\nv_{n}$.
The center coordinate of the system $n$ is represented by $\Xv_n$, and the spacing of the systems is $d$ (i.e., $|\Xv_n-\Xv_m|=d$ for a neighbouring pair.
The electron hopping occurs on the lead (shown by obals), between sites $\Rv_{nm}$ and $\Rv_{mn}$.
\label{FIGseebecksystem}
}
\end{center}
\end{figure}

We model the temperature gradient by considering a discretized model
consisting of the systems labeled by $n=1,2\cdots$.
Each system $n$ is assumed to be in local thermal equilibrium at temperature $T_n$ and chemical potential $\mu_n$ (Fig. \ref{FIGseebecksystem}). 
(In the end, we will take the continuum limit, assumeing that the temperature gradient is not very large. )
Without losing generality, we assume that systems are placed on a cubic lattice with equal distance $d$. 
The conduction electrons in each subsystem are represented by plane waves whose wave vectors $\kv$ are approximated to take any value. 
The magnetization direction of each system, $\nv_n$, is assumed to be uniform within the system but is different for different $n$.
The Hamiltonian of the systems when isolated is given as
\begin{align}
H_0 &= \sum_{n} \sum_{\kv} 
\cdag_{n\kv} \lt(\epsilon_{\kv}-\spol \nv_n\cdot\sigmav\rt) c_{n\kv},
\end{align}
where $\spol$ is the spin spitting energy due to the magnetization.
The electron operator is represented by a two-component field, $c_{n\kv}=(c_{n\kv+}, c_{n\kv-} )$, where 
$\pm$ represents the spin.
To describe the magnetization $\nv_n$ dependent on $n$, a gauge transform in the spin space that diagonalize the exchange interaction is useful.
This transform is carried out as
\begin{align}
c_{n\kv}=U_n a_{n\kv},
\end{align}
where $a_{n\kv}$ is a new electron operator in the gauge-transformed frame and $U_n$ is a $2\times2$ unitary matrix, given as
\begin{align}
U_n & \equiv \mv_n\cdot\sigmav \nnr 
\mv_n &\equiv (\sin\frac{\theta_n}{2}\cos\phi_n,\sin\frac{\theta_n}{2}\sin\phi_n,\cos\frac{\theta_n}{2}),
\end{align}
 with
$(\theta_n,\phi_n)$ being the polar coordinates representing $\nv_n$ \cite{TKS_PR08}.

%

By the gauge transformation, the Hamiltonian of the subsystems when isolated becomes 
\begin{align}
H_0 &= \sum_{n} \sum_{\kv\pm} 
\epsilon_{\kv\pm} \adag_{n\kv\pm}a_{n\kv\pm},
\end{align}
where $\epsilon_{\kv\pm}\equiv \frac{k^2}{2m} \mp\spol$, and $\pm$ is the spin index.

The subsystems are connected by leads, where the electron hopping occurs.
The coordinate in the lead in a system $n$ where the hopping to a neighbouring system $m$ occurs is represented by $\Rv_{nm}$.
The hopping Hamiltonian reads  (in the real space representation) 
\begin{align}
H_t &= \sum_{\average{nm}} \sum_{\Rv_{nm}\Rv_{mn}} \sum_{\pm}
 t  (\cdag_{m\pm}(\Rv_{mn}) c_{n\pm}(\Rv_{nm})+\cdag_{n\pm}(\Rv_{nm}) c_{m\pm}(\Rv_{mn})), \label{Hhop1}
\end{align}
where $\average{nm}$ denotes a pair of neighbouring systems.
After the gauge transform, it reads
\begin{align}
H_t &= \sum_{\average{\L\R}}  \sum_{\Rv_{\R\L}\Rv_{\L\R}}  
 t  (\adag_{\L}(\Rv_{\L\R})U_{\L\R} a_{\R}(\Rv_{\R\L})+\adag_{\R}(\Rv_{\R\L}) U_{\R\L} a_{\L}(\Rv_{\L\R})), \label{Hhop2}
\end{align}
where $U_{\L\R}\equiv U_{\L}^\dagger U_{\R}$.

In terms of $a_\L$ electron, the spin density of the $\L$ system is written as 
$\sv_\L\equiv \average{c_\L^\dagger \sigmav c_\L}
= \average{a_\L^\dagger U_\L^\dagger \sigmav U_\L  a_\L}$.
We represent the charge and spin currents through the junction as $I^0_i$ and $I^{\alpha}_i$ ($i=x,y,z$ and $\alpha=x,y,z$ are the spatial and spin direction, respectively).
In the present junction model, the spin (charge) current at system $\L$ is calculated by estimating the time derivative of the spin (charge) density, which reads
\begin{align}
\dot{s}_\L^\alpha(\Rv_{nm})
& =
it \sum_{m}\sum_{\Rv_{mn}}
\average{
a_\L^\dagger(\Rv_{\L\R}) U_\L^\dagger \sigma^\alpha U_\L U_{\L\R} a_\R(\Rv_{\R\L})
-a_\R^\dagger(\Rv_{\R\L}) U_{\R\L} U_\L^\dagger \sigma^\alpha U_\L a_\L(\Rv_{\L\R})
}.
\end{align}
The current flowing in direction $i$ is thus given by
\begin{align}
I^\alpha_i(\L)
&
=-\frac{et}{2} \sum_{\pm}  (\pm) 
\sum_{\Rv_{\R\L}\Rv_{\L\R}} 
\tr[U_\L^\dagger \sigma^\alpha U_\L 
(U_{\L\R} G_{\R\L}(\Rv_{\R\L},\Rv_{\L\R})
{\color{red} -}
G_{\L\R}(\Rv_{\L\R},\Rv_{\R\L})U_{\R\L})]^<_{\Xv_{\R}=\Xv_{\L}\pm \iv d}   ,
\end{align}
where $\iv$ is a unit vector along $i$-axis and 
$m$ is the label of the system neighbouring the system $n$ (i.e., $\Xv_{\R}=\Xv_{\L}\pm \iv d$), and 
$^<$ denotes the lesser component.
Estimating the Green's functions to the lowest (second) order in $t$, we obtain
\begin{align}
I^\alpha_i(\L)
&= -\frac{et^2}{2}  \sum_{\pm} 
\sum_{\Rv_{\R\L}\Rv_{\L\R}\Rv_{\R\L}'\Rv_{\L\R}'} 
  (\pm) 
\tr[U_\L^\dagger \sigma^\alpha U_\L 
\nnr
& \times 
(U_{\L\R} g_{\R}(\Rv_{\R\L},\Rv_{\R\L}') U_{\R\L} g_{\L}(\Rv_{\L\R},\Rv_{\L\R}')
{-}
g_{\L}(\Rv_{\L\R},\Rv_{\L\R}') U_{\L\R}g_{\R}(\Rv_{\R\L},\Rv_{\R\L}') U_{\R\L} )]^< _{\Xv_{\R}=\Xv_{\L}\pm \iv d}    ,
\end{align}
where $g_\R$ is the free Green's function of the system $\R$ on the Keldysh contour.
Without losing generality, we choose $\nv_\L$ as along $z$ direction, i.e., $U_{\L}=1$.
Since we are eventually interested in the continuum limit, we consider the case where the difference between $\nv_\R$ and $\nv_\L$ is small.
The rotation matrix $U_{\L\R}$ is expressed by a spin gauge field 
$\Av_{\L\R}$ defined as
$U_{\L\R} \equiv e^{i\Av_{\L\R} \cdot \sigmav}
=U_{\R\L}^\dagger$.
Explicitely, $U_{\L\R} = \mv_\L\cdot\mv_\R +i\sigmav\cdot(\mv_\L\times\mv_\R)$ and thus
$\Av_{\L\R}=(\mv_\L\times \mv_\R)$.
The current then reads 
\begin{align}
I^{\alpha}_i(\L)
 &=
\frac{e}{2}\sum_{\pm} (\pm)  \sumom \sum_{\kv\kv'}|t_{\kv'\kv}|^2
\tr[\sigma^\alpha \lt\{
\green{\L\kv\omega}{\ret} e^{i\Av_{\L\R}\cdot\sigmav} 
F_{\R\kvp}(\green{\R\kvp\omega}{\adv}-\green{\R\kvp\omega}{\ret})
e^{-i\Av_{\L\R}\cdot\sigmav}  \rt.  \nnr
& +
F_{\L\kv}
(\green{\L\kv\omega}{\adv}-\green{\L\kv\omega}{\ret})
e^{i\Av_{\L\R}\cdot\sigmav} \green{\R\kvp\omega}{\adv} 
e^{-i\Av_{\L\R}\cdot\sigmav} 
 {-}
 e^{i \Av_{\L\R}\cdot\sigmav} \green{\R\kvp\omega}{\ret}
e^{-i\Av_{\L\R}\cdot\sigmav} F_{\L\kv}(\green{\L\kv\omega}{\adv}-\green{\L\kv\omega}{\ret})
\nnr
& \lt. 
 {-}
 e^{i\Av_{\L\R}\cdot\sigmav} F_{\R\kvp}
(\green{\R\kvp\omega}{\adv}-\green{\R\kvp\omega}{\ret})
e^{-i\Av_{\L\R}\cdot\sigmav} \green{\L\kv\omega}{\adv} 
\rt\} ]_{\Xv_{\R}=\Xv_{\L}\pm \iv d}
\nnr
&=
{2e}\sum_{\pm} (\pm)  \sumom \sum_{\kv\kv'}|t_{\kv'\kv}|^2
\tr[\sigma^\alpha (F_{\R\kvp}-F_{\L\kv})
\Im[\green{\L\kv\omega}{\adv}] 
\Im[\green{\R\kvp\omega}{\adv}]
\nnr
&
-e\sum_{\pm}(\pm)  \sumom \sum_{\kv\kv'}|t_{\kv'\kv}|^2 
A_{\L\R}^\beta
\tr[\sigma^\alpha 
\{\green{\L\kv\omega}{\ret} 
[\sigma^\beta,F_{\R\kvp}\Im(\green{\R\kvp\omega}{\adv})]
  \nnr
& +
F_{\L\kv}
\Im(\green{\L\kv\omega}{\adv})
[\sigma^\beta,\green{\R\kvp\omega}{\adv} ]
{-}
[\sigma^\beta, \green{\R\kvp\omega}{\ret}]
F_{\L\kv}\Im(\green{\L\kv\omega}{\adv})
{-}
 [\sigma^\beta, F_{\R\kvp} \Im(\green{\R\kvp\omega}{\adv})]
\green{\L\kv\omega}{\adv} 
\}]_{\Xv_{\R}=\Xv_{\L}\pm \iv d}
\nnr
&\equiv I_{0,i}^\alpha+\delta I^\alpha_i , \label{I1}
\end{align}
where
\begin{align}
t_{\kv',\kv}\equiv t \sum_{\Rv_{\R\L}\Rv_{\L\R}}
e^{i\kv'\cdot\Rv_{\R\L}}e^{-i\kv\cdot\Rv_{\L\R}},
\end{align}
and 
 $I_0^\alpha$ and $\delta I^\alpha$ represent the contribution without the gauge field and the linear order contribution, respectively.
We neglect the higher order contribution in the gauge field, since we consider a slowly varying spin texture.
The geometry of the lead is reflected in the amplitude $t_{\kv',\kv}$.
The Fermi distribution function is represented by matrix 
\begin{align}
F_{\R\kv}\equiv 
\lt(
  \begin{array}{cc}
    f_{\R}(\epsilon_{\kv+}) & 0 \\
     0 & f_{\R}(\epsilon_{\kv-})  \end{array} \rt),
\end{align}
where ($\beta_{\R}\equiv (\kB T_{\R})^{-1}$)
\begin{align}
f_{\R}(\ekvs)\equiv\frac{1}{e^{\beta_{\R}(\ekvs-\mu_{\R})}}.
\end{align}
Retarded Green's function $\green{\R\kv\omega}{\ret}$ is a $2\times 2$ matrix in spin space with each component defined as
\begin{align}
\green{\R\kv\sigma\omega}{\ret}
 = \frac{1}{\hbar\omega-\ekvs+i\eta},
\end{align}
where 
$\eta$ represents an infinitesimal positive 
(or the inverse lifetime if disordered).

\section{Uniform magnetization}

Let us first consider the contribution $I_0^\alpha$,  
the current when the magnetization of the whole system is uniform.
Explicitely writing the spin index, the currents read  
\begin{align}
I_{0,i}^{0}(\L)
 &=
2e \sum_{\pm}  (\pm) \sumom \sum_{\kv\kv'}|t_{\kv'\kv}|^2
\sum_{\sigma}
(f_\R(\ekvps)-f_\L(\ekvs))\Im[\green{\L\kv\sigma\omega}{\ret}]
\Im[\green{\R\kvp\sigma\omega}{\ret}]_{\Xv_{\R}=\Xv_{\L}\pm \iv d}
\nnr
I_{0,i}^{z}(\L)
 &=
2 e \sum_{\pm}  (\pm) \sumom \sum_{\kv\kv'}|t_{\kv'\kv}|^2
\sum_{\sigma}
\sigma (f_\R(\ekvps)-f_\L(\ekvs))\Im[\green{\L\kv\sigma\omega}{\ret}]
\Im[\green{\R\kvp\sigma\omega}{\ret}]_{\Xv_{\R}=\Xv_{\L}\pm \iv d}, \label{Iz0}
\end{align}
and $I_0^{x}=I_0^{y}=0$.
We consider the case of an infinitesimal difference of the temperature and the chemical potential of the two adjacent systems, $\L$ and $\R$.
Defining $T_\R\equiv T+\Delta_{\R} T$ and $\mu_{\R}=\mu+\Delta_{\R}\mu$ 
($T$ and $\mu$ are the temperature and the chemical potential for the electron of the system $\L$), we expand the physical quantities up to the linear order in the differences.
 The difference of the Fermi distribution functions for $\L$ and $\R$ is written as
\begin{align}
(f_\R(\ekvps)-f_\L(\ekvs))
&=(f_\L(\ekvps)-f_\L(\ekvs))
+ f'(\ekvps)
\lt(\Delta_{\R} \mu+(\ekvps-\mu)\frac{\Delta_{\R} T}{T}\rt),
\end{align}
where 
\begin{align}
f'(\ekvps)
= -\frac{\beta}{4} \frac{1}{\cosh^2\frac{\beta}{2}(\ekvps-\mu)}.
\end{align}

Let us first consider the conventional Seebeck effect, i.e., the charge part. 
The charge current reads
\begin{align}
I_{0,i}^{0}
 &=
-e \eta^2\frac{\beta}{2} 
\sumom \sum_{\kv\kv'}|t_{\kv'\kv}|^2
\sum_{\sigma}
\frac{1}{(\omega-\ekvs)^2+\eta^2}
\frac{1}{(\omega-\ekvps)^2+\eta^2}
\frac{1}{\cosh^2\frac{\beta}{2}(\ekvps-\mu)}\nnr
& \times 
\sum_{\pm} (\pm)
\lt(\Delta_{\R}\mu+(\ekvps-\mu)\frac{\Delta_{\R} T}{T}\rt)_{\Xv_{\R}=\Xv_{\L}\pm \iv d} .
\end{align}
The $\omega$-integration is carried out as
\begin{align}
\eta^2  \sumom 
\frac{1}{(\omega-\ekvs)^2+\eta^2}
\frac{1}{(\omega-\ekvps)^2+\eta^2}
&=
\frac{\eta}{2}\frac{1}{(\ekvs-\ekvps)}
\lt(\frac{1}{(\ekvs-\ekvps+2i\eta)}
+{\rm c.c.}\rt) \nnr
&=\frac{\eta}{(\ekvs-\ekvps)^2+4\eta^2} ,
\end{align}
and  thus
\begin{align}
I_{0,i}^{0}
 &=
-e \eta  \frac{\beta}{2}  
\sum_{\kv\kv'}|t_{\kv'\kv}|^2
\sum_{\sigma}
\frac{1}{(\ekvs-\ekvps)^2+4\eta^2}
\frac{1}{\cosh^2\frac{\beta}{2}(\ekvps-\mu)}
\nnr
& \times 
\sum_{\pm} (\pm)
\lt(\Delta_{\R}\mu+(\ekvps-\mu)\frac{\Delta_{\R} T}{T}\rt)_{\Xv_{\R}=\Xv_{\L}\pm \iv d} .
\end{align}
We consider now the continuum limit by taking $d\ra 0$.
This is allowed when $d \ll \Lambda$, where $\Lambda$ is the spatial length scale the temperature and the chemical potential varies significantly , i.e., 
$\Lambda\simeq O\lt(\lt(\frac{\nabla T}{T}\rt)^{-1},\lt(\frac{\nabla \mu}{\mu}\rt)^{-1}\rt)$.
The summation over $\R$ is carried out easily as (choosing $\Xv_{\L}=0$)
\begin{align}
\hf\sum_{\pm}(\pm) \Delta_{\R}\mu|_{\Xv_{\R}=\Xv_{\L}\pm \iv d}
&= \hf[\mu(+d\iv)-\mu(0)-(\mu(-d\iv)-\mu(0))]
= d \nabla_i \mu.
\end{align}
We thus see that  
\begin{align}
\Iv^{0}_{0}
 &= -d(
G \nabla \mu/e+G_T\nabla T), \label{I0G}
\end{align}
where the conductance $G$ and the thermal conductance $G_T$ as functions of the chemical potential are given as
\begin{align}
G  (\mu) &=
2 e^2  
\sum_{\kv\kv'}
\sum_{\sigma}
\frac{\eta  |t_{\kv'\kv}|^2}{(\ekvs-\ekvps)^2+4\eta^2}
\frac{\beta/4 }{\cosh^2\frac{\beta}{2}(\ekvps-\mu)}
\nnr
G_T (\mu) &=
2 e^2  
\sum_{\kv\kv'}
\sum_{\sigma}
\frac{\eta  |t_{\kv'\kv}|^2}{(\ekvs-\ekvps)^2+4\eta^2}
\frac{\kB \beta^2(\ekvps-\mu)/4}{\cosh^2\frac{\beta}{2}(\ekvps-\mu)},
\end{align}
respectively.

The conductance is written by use of the Boltzmann conductivity $\sigma_{\rm B}$  as $G=\sigma_{\rm B} A/d$ where $A$ is the area of each system.
Let us here switch to the current density, defined as $j\equiv I/A$.
Then \Eqref{I0G} reduces to 
\begin{align}
\jv_0 &=
-( \sigma_{\rm B} \nabla \mu/e +\sigma_{T}\nabla T). \label{jsigma}
\end{align}
The conductivity is given in terms of the spin resolved conductivity  $\sigma_{{\rm B},\sigma}$ ($\sigma=\pm$ is the spin index) as 
\begin{align}
\sigma_{{\rm B}} &=
\sum_\sigma \int_{-\sigma\spol}^{\infty} d\epsilon \sigma_{{\rm B},\sigma}(\epsilon) 
\frac{\beta/4 }{\cosh^2\frac{\beta}{2}(\epsilon-\mu)}.
\end{align}
where 
\begin{align}
\sigma_{{\rm B},\sigma}(\epsilon) 
&\equiv 
2e^2\frac{d}{A} \sum_{\kv\kv'}
\frac{\eta  |t_{\kv'\kv}|^2}{(\ekvs-\ekvps)^2+4\eta^2}
\delta(\epsilon-\ekvps). \label{conductivity}
\end{align}
At low temperatures, $\frac{\beta/4 }{\cosh^2\frac{\beta}{2}(\epsilon-\mu)}
=\delta(\epsilon-\mu)$ and we reproduce 
$\sigma_{{\rm B}}=\sum_{\sigma}\sigma_{{\rm B},\sigma}$.
Defining 
$x\equiv \frac{\beta}{2}(\epsilon-\mu)$, the thermal conductivity is written as
\begin{align}
\sigma_T
&= \frac{\kb}{e} \sum_{\sigma=\pm} \int_{-\frac{\beta}{2}(\mu+\sigma\spol)}^\infty dx  \frac{x}
{\cosh^2x}
\sigma_{{\rm B},\sigma}|_{\epsilon=\mu+\frac{2x}{\beta}} .
\end{align}
At low temperature, $\beta\mu\gg1$, we can expand the integrand with respect to $x$ to obtain the well-known relation 
\begin{align}
\sigma_T
&= 
 \frac{\kb^2 T}{e} \frac{\pi^2}{3} \sum_{\sigma}\lt.\frac{d\sigma_{{\rm B},\sigma}}{d\epsilon}\rt|_{\epsilon=\mu}
=
\sigma_{{\rm B}} \times \frac{\kb^2 T}{e}  \frac{\pi^2}{3} \frac{\sum_{\sigma}
\lt.\frac{d\sigma_{{\rm B},\sigma}}{d\epsilon}\rt|_{\epsilon=\mu}
}{\sum_{\sigma}\sigma_{{\rm B},\sigma}(\mu)}.\label{sigmaTderivation}
\end{align}

From the above result, the spin current density, $j_0^z\equiv I_0^z /A$, is easily obtained as 
$j_0^{z}
 =
\sigma_s \nabla \mu/e +\sigma_{s,T}\nabla T
$, 
where 
\begin{align}
\sigma_s &=
\sum_{\sigma=\pm} \sigma \int_{-\sigma\spol}^{\infty} d\epsilon \sigma_{{\rm B},\sigma}(\epsilon) 
\frac{\beta/4 }{\cosh^2\frac{\beta}{2}(\epsilon-\mu)},
\end{align}
and
\begin{align}
\sigma_{s,T}
&=
\sigma_s \times \frac{\kb^2 T}{e} \frac{\pi^2}{3} \frac{\sum_{\sigma}\sigma 
\lt.\frac{d\sigma_{{\rm B},\sigma}}{d\epsilon}\rt|_{\epsilon=\mu}
}
{\sum_{\sigma}\sigma\sigma_{{\rm B},\sigma}(\mu)}.
\end{align}
We therefore see that the temperature gradient acts on the spin current as an effective electric field in agreement with naive guess.
When the magnetization is along $\nv$, the above result of spin current becomes 
\begin{align}
j_0^{\alpha}
 &=
-n^\alpha( \sigma_s \nabla \mu/e +\sigma_{s,T}\nabla T). \label{jssigma}
\end{align}

The Seebeck coefficient is defined as the ratio 
\begin{align}
S
&\equiv \frac{\sigma_T}{\sigma_{{\rm B}}}
= \frac{\kb^2 T}{e}  \frac{\pi^2}{3} \frac{\sum_{\sigma}
\lt.\frac{d\sigma_{{\rm B},\sigma}}{d\epsilon}\rt|_{\epsilon=\mu}
}{\sum_{\sigma}\sigma_{{\rm B},\sigma}(\mu)}.
\end{align}
We define the spin Seebeck coefficient as
\begin{align}
S_s
&\equiv \frac{\sigma_{s,T}}{\sigma_s}
= \frac{\kb^2 T}{e}  \frac{\pi^2}{3} \frac{\sum_{\sigma}\sigma
 \lt.\frac{d\sigma_{{\rm B},\sigma}}{d\epsilon}\rt|_{\epsilon=\mu}
}{\sum_{\sigma}\sigma \sigma_{{\rm B},\sigma}(\mu)}.  \label{Ssresult}
\end{align}

Therefore our model reproduces the relations (\ref{Seebeck0})(\ref{spinSeebeck0}) obtained by classical argument \cite{Ashcroft76}.

\section{Examples}
The explicite expressions of $\sigma_{{\rm B},\sigma}$ , $S$ and $S_s$ depend on the 
detail of the hopping on the lead. 
We here present results for three typical cases. 

\subsection{Point-like lead}
We first consider a case of point-like lead.
The coefficient $t_{\kv'\kv}$ then becomes a constant,
$t_{\kv'\kv}=t$ (since we can choose $\Rv_{\L\R}=\Rv_{\R\L}=0$), 
and the summations over $\kv'$ and $\kv$ become independent.
In this case,
\begin{align}
\sigma_{{\rm B},\sigma}(\epsilon) 
&=
2\pi t^2 e^2\frac{d}{A} \sum_{\kv'}
\dos_0 \sqrt{\epsilon_{\kv'}}
\delta(\epsilon-\ekvps)
=
2\pi t^2 e^2\frac{d}{A} 
(\dos_0)^2 (\epsilon+\sigma\spol)
,
\end{align}
where 
$\dos_0\equiv \frac{V m^{3/2}}{\sqrt{2} \pi^2}$ is the three-dimensional density of states divided by $\sqrt{\epsilon}$ ($\epsilon$ is the enery and $V=Ad$).
(We note that in taking the pointlike limit, $t^2d(\dos_0)^2/A$ needs to be kept a constant, since the combination $t^2d(\dos_0)^2/A$ gives the physical conductance as seen in the above equation.)
We then obtain 
\begin{align}
\frac{\sigma_T}{\sigma_{\rm B}}
&= \frac{\pi^2}{3}\frac{\kb^2T}{e\mu},
\end{align}
while 
\begin{align}
\sigma_{s,T}=0,
\end{align}
since $\frac{d\sigma_{{\rm B},\sigma}(\epsilon)}{d\epsilon}$ is a constant independent of the spin. 
Thus spin Seebeck coefficient vanishes if the lead is point-like and if the conduction electron's energy is the free electron type, $k^2$.

\subsection{Two-dimensional interfaces}
If the junction in the discretized model is a thin plane with electron scattering, the electron hopping conserves the wave vector perpendicular to the junction (which we denote $\kv_\perp$) but not the component along the junction  (we choose the junction along $x$ axis).
The wave vectors before and after the hopping are thus written as
$\kv=(k,\kv_\perp)$ and $\kv'=(k',\kv_\perp)$, respectively, where $k$ and $k'$ are independent.
When the continuum limit is taken in this interface model, the summation over the wave vectors in \Eqref{conductivity} is carried out to obtain 
\begin{align}
\sigma_{{\rm B},\sigma}(\epsilon) 
&=
2\pi t^2 e^2\frac{d}{A} 
(\dos^{(1)}_0)^2 \dos^{(2)}_0 \sqrt{\epsilon+\sigma\spol}
,
\end{align}
where $\dos^{(1)}_0\equiv \frac{\sqrt{m}d}{2\sqrt{2} \pi}$ and $\dos^{(2)}_0\equiv \frac{mA}{2\pi}$
are  the coefficients in the one- and two-dimensional density of states, respectively.
Therefore
\begin{align}
\frac{\sigma_{s,T}}{\sigma_{s}}
&= -\frac{\pi^2}{6}\frac{\kb^2T}{e\sqrt{\mu^2-\spol^2}}
= -\frac{\sigma_{T}}{\sigma_{\rm B}}
.
\end{align}

\subsection{Free electron limit}
Our model can also describe the free electron limit.
To describe the free electron with mass $m$, the parameter $t$ and $t_{kk'}$ of the tight-binding Hamiltonian need to be replaced by
$t\simeq \frac{1}{md^2}$
(using $t\cos(k_xd)=\frac{k_x^2}{2m}+{\rm const.}$)
and 
$t_{kk'}=t k_x d\delta_{kk'}$ (i.e., $t_{kk'}^2=\frac{k_x^2}{m^2d^2} \delta_{kk'}$), respectively. 
(We choose the transport as along $x$ direction.)
We then reproduce from \Eqref{conductivity} the Boltzmann conducitivity 
\begin{align}
\sigma_{{\rm B},\sigma}(\epsilon) 
&=
\frac{e^2 n_\sigma(\epsilon)\tau}
{m^2},
\end{align}
where $n_\sigma(\epsilon)\equiv \frac{[k(\epsilon+\sigma\spol)]^3}{6\pi^2}$ 
($k(\epsilon+\sigma\spol)=\sqrt{2m(\epsilon+\sigma\spol)}$).
The spin Seebeck coefficient then reads (from \Eqref{Ssresult})
\begin{align}
S_s
&= \frac{\kb^2 T}{e}  \frac{\pi^2}{2} 
\frac
{(\mu+\spol)^{\frac{1}{2}} -(\mu-\spol)^{\frac{1}{2}}}
{(\mu+\spol)^{\frac{3}{2}} -(\mu-\spol)^{\frac{3}{2}}}.
\end{align}

\section{Gauge field (spin texture) contribution}

We consider here the contribution linear in the gauge field, 
$ \delta I_i^\alpha$, in the adiabatic limit, namely,  
the spin variation is assumed to be small so that the momentum transfer due to the gauge field is neglected compared with the conduction electron's momentum. 
The linear contribution in \Eqref{I1}  reads 
\begin{align}
\delta I^{\alpha}_i(\L)
 &=
-e\sum_{\pm}(\pm) \sumom \sum_{\kv\kv'}|t_{\kv'\kv}|^2 
A_{\L\R}^\beta
\tr[\sigma^\alpha 
\{\green{\L\kv\omega}{\ret} 
[\sigma^\beta,F_{\R\kvp}\Im(\green{\R\kvp\omega}{\adv})]
  \nnr
& +
F_{\L\kv}
\Im(\green{\L\kv\omega}{\adv})
[\sigma^\beta,\green{\R\kvp\omega}{\adv} ]
{-}
[\sigma^\beta, \green{\R\kvp\omega}{\ret}]
F_{\L\kv}\Im(\green{\L\kv\omega}{\adv})
{-}
 [\sigma^\beta, F_{\R\kvp} \Im(\green{\R\kvp\omega}{\adv})]
\green{\L\kv\omega}{\adv} 
\}]_{\Xv_{\R}=\Xv_{\L}\pm \iv d}
\nnr
&=
-e\sum_{\pm}(\pm)   \sumom \sum_{\kv\kv'}|t_{\kv'\kv}|^2 
A_{\L\R}^\beta
\tr[ 
( \sigma^\alpha \green{\L\kv\omega}{\ret}
 - \green{\L\kv\omega}{\adv} \sigma^\alpha ) 
[\sigma^\beta,F_{\R\kvp}\Im(\green{\R\kvp\omega}{\adv})]
\nnr
& 
+ \sigma^\alpha[ F_{\L\kv}\Im(\green{\L\kv\omega}{\adv})
[\sigma^\beta,\green{\R\kvp\omega}{\adv} ]
- 
[\sigma^\beta,\green{\R\kvp\omega}{\ret} ]
F_{\L\kv}\Im(\green{\L\kv\omega}{\adv})
]]_{\Xv_{\R}=\Xv_{\L}\pm \iv d} 
. \label{I2}
\end{align}
By use of 
$[\sigma^\alpha, \green{\L\kv\omega}{\ret}] 
=-2i \sum_{\gamma} \epsilon_{\alpha\gamma z}\sigma^\gamma \sum_{\sigma} \sigma \green{\L\kv\omega\sigma}{\ret}$, 
the spin part ($\alpha=x,y,z$) reads  
\begin{align}
\delta I^{\alpha}_i (\L)
 &=
A \sum_{\beta} 
[a_{i,\beta} \epsilon_{\alpha\beta z}
+(b^{\rm (eq)}_{i,\beta} +b^{\rm (d)}_{i,\beta})
 (\delta_{\alpha\beta}-\delta_{\alpha,z}\delta_{\beta,z}) ], \label{deltaI1}
\end{align}
where $A$ is the junction area and the coefficients are 
\begin{align}
a_{i,\beta} &=
-\frac{2e}{A}\sum_{\pm}(\pm) \sumom \sum_{\kv\kv'}|t_{\kv'\kv}|^2 A_{\L\R}^\beta 
\sum_{\sigma\sigma'}  \sigma' (f_{\L\kv\sigma}- f_{\R\kvp\sigma'})
\Im (\green{\L\kv\omega\sigma}{\adv})
\Im (\green{\R\kvp\omega\sigma'}{\adv})_{\Xv_{\R}=\Xv_{\L}\pm \iv d} 
 \nnr
b^{\rm (eq)} _{i,\beta}
&=
\frac{2e}{A} \sum_{\pm}(\pm) \sumom \sum_{\kv\kv'}|t_{\kv'\kv}|^2 A_{\L\R}^\beta 
\sum_{\sigma\sigma'} \sigma \sigma'
f_{\L\kv\sigma}
\Im(\green{\L\kv\omega\sigma}{\adv} \green{\R\kvp\omega\sigma'}{\adv}) _{\Xv_{\R}=\Xv_{\L}\pm \iv d} 
\nnr
b^{\rm (d)} _{i,\beta}
&=
-\frac{2e}{A} \sum_{\pm}(\pm) \sumom \sum_{\kv\kv'}|t_{\kv'\kv}|^2 A_{\L\R}^\beta 
\sum_{\sigma\sigma'} \sigma \sigma'
(f_{\L\kv\sigma}-f_{\R\kvp\sigma'}) 
\Re (\green{\L\kv\omega\sigma}{\adv}) \Im(\green{\R\kvp\omega\sigma'}{\adv}) _{\Xv_{\R}=\Xv_{\L}\pm \iv d} 
 \nnr
. \label{I3}
\end{align}
The coefficient $b^{\rm (eq)}$ represents the equilibrium spin current 
and $b^{\rm (d)}$ represents the driven contribution.

By writing 
$f_{\R\kvp\sigma'}=f_{\kvp\sigma'}+\delta f_{\kvp\sigma'}$
($f_{\kvp\sigma'}\equiv f_{\L\kvp\sigma'}$ and 
$\delta f_{\kvp\sigma'}\equiv
 f'(\epsilon_{\kvp\sigma'})
\lt(\Delta_{\R} \mu+(\epsilon_{\kvp\sigma'}-\mu)\frac{\Delta_{\R} T}{T}\rt)$),
the summation over the spatial directions is carried out by expanding the gauge field and the chemical potential 
as (we choose $\Xv_{\L}=0$) 
\begin{align}
\hf\sum_{\pm} 
(\pm) A_{\L\R}^\beta \Delta_{\R} \mu|_{\Xv_{\R}=\Xv_{\L}\pm \iv d}
&=
\hf\sum_{\pm}(\pm) 
(\mv(0)\times \mv(\pm d\iv))^\beta(\mu(\pm d\iv)-\mu(0))
\nnr
&=
{d^3}(\nabla_i^2 \mu)A_i^\beta+o(d^4,\nabla A ),
\end{align} 
where $A_i^\beta\equiv \hf(\mv\times \nabla_i\mv)^\beta$ is the spin gauge field in the continuum limit \cite{TKS_PR08}.
We have neglected the contribution containing the derivative of $A_i^\beta$, since it corresponds to the second order contribution with respect to the gauge field, which we neglect in $\delta I$.
The equilibrium and the linear contributions are given as (we here suppress the index $\L$ in the Green's functions)
\begin{align}
a_{i,\beta} &=
2 A_i^\beta (a_E (\nabla_i)^2 \mu/e  + a_T (\nabla_i)^2 T ) \nnr
b^{\rm (d)} _{i,\beta}
&=
2 A_i^\beta (b_E (\nabla_i)^2 \mu/e  + b_T (\nabla_i)^2 T )  \nnr
b^{\rm (eq)} _{i,\beta}
&= 2 A_i^\beta j^{\rm (eq)}  ,
 \label{I4}
\end{align}
where
\begin{align}
a_E & = \sum_{\sigma} \sigma a_{\sigma} \nnr
a_T & = \lt. 
 \frac{\pi^2}{3} \frac{\kb^2 T}{e} \sum_{\sigma} \sigma \frac{d a_{\sigma}}{d\epsilon} \rt|_{\epsilon=\mu} \nnr
a_{\sigma}(\epsilon) & \equiv 
  -2e \frac{d^3}{A} \sum_{\kv\kv'\sigma'} \sigma'
\frac{\eta  |t_{\kv'\kv}|^2}{(\ekvs-\epsilon_{\kv\sigma'})^2+4\eta^2}
\delta(\epsilon-\ekvs)   \nnr
b_E & = \sum_{\sigma} \sigma b_{\sigma} \nnr
b_T & = \lt. 
 \frac{\pi^2}{3} \frac{\kb^2 T}{e} \sum_{\sigma} \sigma \frac{d b_{\sigma}}{d\epsilon} \rt|_{\epsilon=\mu} \nnr
b_{\sigma}(\epsilon) & \equiv 
  -2e \frac{d^3}{A} \sum_{\kv\kv' \sigma'} \sigma' 
\frac{  |t_{\kv'\kv}|^2 (\ekvs-\epsilon_{\kv\sigma'}) }{(\ekvs-\epsilon_{\kv\sigma'})^2+4\eta^2}
\delta(\epsilon-\ekvs)   ,
\end{align}
and 
\begin{align}
j^{\rm (eq)} 
&=  2e \frac{d^3}{A}  \sumom \sum_{\kv\kv'\sigma \sigma'} \sigma\sigma' 
|t_{\kv'\kv}|^2 f_{\kv\sigma} \Im( \green{\kv\omega\sigma}{\adv} \green{\kv'\omega\sigma'}{\adv} ),
\end{align}
represents the equilibrium current.

The result (\ref{deltaI1}) is for the magnetization at $\L$ is along the $z$ axis (i.e., $U_\L=1$). 
A general case with magnetization along $\nv$ is obtained by applying an unitary transformation
defined by a matrix $V_{\alpha\beta}=2m_\alpha m_\beta-\delta_{\alpha\beta}$.
By use of the identities \cite{TKS_PR08} 
\begin{align}
(2m_\alpha m_\beta-\delta_{\alpha\beta})(\delta_{\beta\gamma}-\delta_{\beta z}\delta_{\gamma z})A_i^\gamma 
&=-\hf(\nv\times \nabla_i\nv)^\alpha 
\nnr
(2m_\alpha m_\beta-\delta_{\alpha\beta})\epsilon_{\beta\gamma z} A_i^\gamma
&=-\hf \nabla_i\nv^\alpha,
\end{align} 
the final result of the gauge field contribution to the spin current density ($\delta j^{\alpha}_i\equiv \delta I^{\alpha}_i/A$) is given by
\begin{align}
\delta j^{\alpha}_i
 & =  -[
 (\nv\times \nabla_i\nv)^\alpha
j^{{\rm (eq)}} 
+ (\nabla_i \nv)^\alpha
(a_E  \nabla_i^2 \mu/e + a_{T} \nabla_i^2 T ) \nnr
& 
+ (\nv\times \nabla_i \nv)^\alpha
(b_E  \nabla_i^2 \mu/e + b_{T} \nabla_i^2 T ).
\label{deltaj}
\end{align}

To summarize the results of the spin current, 
the total charge current in the system is given by Eq. (\ref{jsigma}) as
\begin{align}
j_i&=j_{0,i}\nnr
&= -(\sigma_{\rm B} \nabla_i \mu/e +\sigma_{T}\nabla_i T),
\end{align}
and the spin current reads (Eqs. (\ref{jssigma})(\ref{deltaj}))
\begin{align}
j_{{\rm s},i}^{\alpha} 
& \equiv  j_{0,i}^{\alpha}+ \delta j^{\alpha}_i \nnr
&=
\nv^\alpha (\sigma_s \nabla_i \mu/e+\sigma_{s,T}\nabla_i T)
-
(\nabla_i\nv)^\alpha
( a_E  \nabla_i^2 \mu/e +a_T \nabla_i^2 T )
\nnr
& 
-
(\nv\times \nabla_i\nv)^\alpha
( j^{{\rm (eq)}} + b_E  \nabla_i^2 \mu/e +b_T \nabla_i^2 T )
.
\label{jstotalresult}
\end{align}
The spin current driven by the field and the spin texture (spin gauge field) was calculated here by assuming adiabatic condition. This is justified 
when the electron mean free path $\ell$ satisfies $\ell \gg d$, namely, 
either in the strongly disordered or in the weak hopping regime where $\ell \propto 1/t^2$ is large.

\section{Spin relaxation torque}

In this section, we calculate the spin relaxation torque induced by the temperature gradient by including the spin-orbit interaction.
We consider an uniform magnetization case and neglect the spin gauge field.
The leads connecting subsystems are assumed here to be point-like for simplicity, namely, $t_{\kv'\kv}=t$.
The spin-orbit interaction is 
\begin{align}
\Hso &=  -\frac{i}{2}
\sum_{ijk} \epsilon_{ijk}\intx  (\nabla_i \Vso^{(k)}) 
(\cdag \vvec{\nabla}_j \sigma_k c).
\end{align}
The spin-orbit potential $\Vso^{(k)}$ is assumed to arise from random impurities and to depend on the spin direction ($k$).
The impurity scattering is treated in the standard manner \cite{TKS_PR08}.

By  deriving the continuity equation for the spin density, the dominant spin relaxation torque in $z$ direction 
acting in the system $\L$ is found to be \cite{Nakabayashi10}
\begin{align}
{\cal T}^{z} (\L)
&\equiv 
i \sum_{ijkl} \epsilon_{ijk} \epsilon_{z lk} (\nabla_i\Vso^{(k)})
\average{ \cdag_{\L} \sigma_l \vvec{\nabla}_j c_{\L}}.
\label{Tsodef} 
\end{align}
It is calculated including the hopping to other subsystems at the second order as
\begin{align}
{\cal T}^z(\L) & = 
 - \frac{4}{9} i \nso \lamso^2 t^2 \sum_{\beta\gamma} \sum_{\R}
\sum_{\kv\kv'\kv''} \sumom k^2  (k'')^2 
\epsilon_{z\beta\gamma}
\tr[ \sigma^\gamma \green{\L\kv''\omega}{} \sigma^\beta
\green{\L\kv\omega}{} \green{\R\kv'\omega}{} 
\green{\L\kv\omega}{} ]^<.
\end{align}
%
\begin{figure}[tbh]
\begin{center}
\includegraphics[width=0.5\hsize]{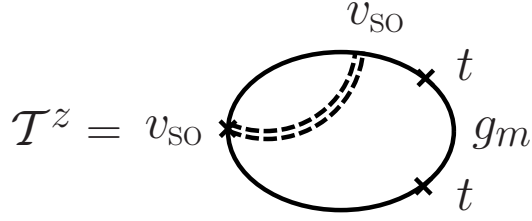}
\caption{ 
The Feynman diagram describing the dominant contribution to the spin relaxation torque induced by the temperature gradient. 
$v_{\rm so}$ represents the spin-orbit interaction, and 
$t$ denotes the electron hopping to other subsystems $\R$ with different temperature $T_\R$.
\label{FIGseebeckTau}
}
\end{center}
\end{figure}
%

Taking the lesser component, we obtain
\begin{align}
{\cal T}^z (\L) & = 
 - \frac{16}{9} i \nso \lamso^2 t^2  \sum_{\beta\gamma} \sum_{\R}
\sum_{\kv\kv'\kv''} \sumom k^2  (k'')^2 
\epsilon_{z\beta\gamma}\sum_{\sigma}  \sigma  \tau_\sigma 
\nnr&\times
\Im(\green{\L\kv'',-\sigma,\omega}{\ret}) (f_\L(\ekvs)-f_\R(\ekvps))
\Im(\green{\L\kv\sigma\omega}{\ret}) 
\Im(\green{\R\kv'\sigma\omega}{\ret}),
\end{align}
where the distribution function of subsystem $n$ is given by
($\beta_{\R}\equiv (\kB T_{\R})^{-1}$)
\begin{align}
f_{\R}(\ekvs)\equiv\frac{1}{e^{\beta_{\R}(\ekvs-\mu_{\R})}+1}.
\end{align}
The summation over the wave vectors are calculated by use of contour integrals, and we obtain
\begin{align}
{\cal T}^z(\L)
& =
 - \frac{32}{9} \pi^2  m^2 \nso \lamso^2 t^2  \dos_0^3   \sum_{\R}
\sum_{\sigma} \sigma \tau_\sigma 
\int_{-\frac{\beta}{2}(\mu-\spol)}^\infty dx
\frac{ 1 }{\cosh^2 x} 
\lt( \Delta\mu_\R +2\kb\Delta T_\R x \rt)
\nnr & \times
\lt(\mu+\sigma\spol+\frac{2x}{\beta}\rt)^{2}
\lt(\mu-\sigma\spol+\frac{2x}{\beta}\rt)^{3/2}
\label{Tau3}
\end{align}
where
$\Delta\mu_\R \equiv \mu_\R-\mu$,
$\Delta T_\R\equiv T_\R-T$ ($T$ and $\mu$ are the temperature and the chemical potential of the system $\L$), 
$x\equiv\frac{\beta}{2} (\ekvp-\sigma\spol-\mu)$.
Considering low temperatures, i.e., $\beta(\mu-\spol)\gg 1$, we can expand the integrand with respect to $x$ and obtain 
\begin{align}
{\cal T}^z(\L)
& = 
 - \frac{64}{9} \pi^2  m^2 \nso \lamso^2 t^2  \dos_0^3   \sum_{\R}
\sum_{\sigma} \sigma \tau_{\sigma} 
\lt(\mu^2-\spol^2\rt)^{2}
\lt(\mu-\sigma \spol\rt)^{-1/2}
\nnr
& \times 
\lt(\Delta\mu_\R
+\frac{\pi^2}{6}\kb^2 T \Delta T_\R 
\frac{7\mu-\sigma\spol}
{\lt(\mu^2-\spol^2\rt)} \rt)
\label{Tau4}
\end{align}

This is the result of a discrete model. 
We now take the continuum limit by replacing
$\Delta \mu_\R $ with  
$\Xv_\R \cdot\nabla \mu
+\sum_{ij}\frac{\Xv_{\R,i}\Xv_{\R,j}}{2} \nabla_i \nabla_j \mu$, and similar expression for $\Delta T_\R$
(we have chosen $\Xv_\L=0$).
Considering the rotationally symmetric system with equal separation $d$ between the small local equilibrium subsystems, $\Xv_\R \equiv \pm d\hat{\rv}$, where $\hat{\rv}$ represents three unit vectors in the three spatial directions,
we obtain 
$\sum_{\R}\Delta\mu_\R=d^2 \nabla^2 \mu$.
The relaxation torque is therefore obtained as
\begin{align}
{\cal T}^z
=\gamma_{E} \nabla^2 \mu/e + \gamma_{T} \nabla^2 T,
\label{Tsores2}
\end{align}
where 
\begin{align}
\gamma_{E} 
& \equiv 
 - \frac{64e}{9} \pi^2  m^2 \nso \lamso^2 t^2  \dos_0^3  d^2  
\lt(\mu^2-\spol^2\rt)^{2}
\sum_{\sigma} \sigma \tau_{\sigma}
\lt(\mu-\sigma \spol\rt)^{-1/2}
\label{gammamu} \\
\gamma_{T}
&=
 - \frac{32}{27} \pi^4 \kb^2 T  m^2 \nso \lamso^2 t^2  \dos_0^3  d^2 
\lt(\mu^2-\spol^2\rt)
\sum_{\sigma} \sigma \tau_{\sigma}
\lt(\mu-\sigma \spol\rt)^{-1/2}
(7\mu-\sigma\spol)
\label{gammaT}
\end{align}
The relaxation torque arises thus from the second order derivatives of $\mu$ and $T$.
The result for $\mu$ here confirmes the result of Ref. \cite{Nakabayashi10} in a discretized model.
(Unlike Ref. \cite{Nakabayashi10}, \Eqref{Tsores2} is symmetric with respect to the spatial direction, but this would be an artifact of the present model, which assumes that the electron hopping occurs on pointlike leads.)
We see that, as is expected, the temperature gradient $\nabla T$ is  equivalent to the electric field in the context of the spin relaxation.

Equation (\ref{gammaT}) indicates that when an uniform temperature gradient is applied to a ferromagnet, the spin relaxation torque is zero. 
Therefore, the spin current driven by homogeneous temperature gradient in the spin Seebeck system  \cite{Uchida08} is spatially uniform without decay, and would be consistent with the experimental observation of the inverse spin Hall signal over the sample of millimeter size. 
We stress here that a term proportional to the spatial coordinate introduced without ground in Ref. \cite{Uchida09} does not exist in the transport equation.

For understanding the experimental result of the thermally induced inverse spin Hall effect, the present analysis needs to be extended to incorporate the spin-charge conversion due to the spin-orbit interaction, which will be carried out in the forthcoming paper.

\section{Conclusion}

To conclude, we have studied the spin current and the spin relaxation torque driven by the temperature gradient microscopically by considering a continuum limit of a discretized model. 
We have shown that the temperature gradient acts as the effective electric field and drives spin current. 
In the uniform magnetization case, 
the spin Seebeck coefficient is given by
\begin{align}
S_s
&= \frac{\kb^2 T}{e}  \frac{\pi^2}{3} 
\frac{\sum_{\pm}(\pm)
 \lt.\frac{d\sigma_{{\rm B},\pm}}{d\epsilon}\rt|_{\epsilon=\mu}
}{\sum_{\pm}(\pm) \sigma_{{\rm B},\pm}(\mu)},
\end{align}
where $\sigma_{{\rm B},\pm}$ is the Boltzamnn conductivity for the spin $\pm$ electron and $\mu$ is the chemical potential.

When the magnetization $\nv$ is nonuniform, spin current components polarized along $\nv\times \nabla \nv$ and 
$\nabla \nv$ are induced by the temperature gradient (\Eqref{jstotalresult}). 
We have also calculated the spin relaxation torque and found that it is proportional to $\nabla^2 T$.
Since the relaxation torque induced by the electric field has been shown to be proportional to $\nabla\cdot \Ev$, 
we see that the temperature gradient $\nabla T$ acts as the effective electric field in the context of the relaxation torque, too.

We have thus demonstrated the equivallence of the temperature gradient and the electric field 
in the spin transport. 
We note, however, that quantitatively these two fields lead to different results since the ratio of the coefficients such as 
$\sigma_T/\sigma_{{\rm B}}$ and $\gamma_T/\gamma_E$ are not always equal.

\acknowledgements
The authors thank K. Uchida, E. Saitoh, K. Taguchi, N. Nakabayashi, G.E.W. Bauer and Y. Tserkovnyak for valuable discussion. 
This work was supported by a Grant-in-Aid for Scientific Research in Priority Areas, "Creation and control of spin current" (1948027), the Kurata Memorial Hitachi Science and Technology Foundation and the Sumitomo Foundation.


\begin{thebibliography}{19}
\expandafter\ifx\csname natexlab\endcsname\relax\def\natexlab#1{#1}\fi
\expandafter\ifx\csname bibnamefont\endcsname\relax
  \def\bibnamefont#1{#1}\fi
\expandafter\ifx\csname bibfnamefont\endcsname\relax
  \def\bibfnamefont#1{#1}\fi
\expandafter\ifx\csname citenamefont\endcsname\relax
  \def\citenamefont#1{#1}\fi
\expandafter\ifx\csname url\endcsname\relax
  \def\url#1{\texttt{#1}}\fi
\expandafter\ifx\csname urlprefix\endcsname\relax\def\urlprefix{URL }\fi
\providecommand{\bibinfo}[2]{#2}
\providecommand{\eprint}[2][]{\url{#2}}

\bibitem[{\citenamefont{Ashcroft and Mermin}(1976)}]{Ashcroft76}
\bibinfo{author}{\bibfnamefont{N.~W.} \bibnamefont{Ashcroft}} \bibnamefont{and}
  \bibinfo{author}{\bibfnamefont{N.}~\bibnamefont{Mermin}},
  \emph{\bibinfo{title}{Solid State Physics}} (\bibinfo{publisher}{Thomson
  Learning}, \bibinfo{year}{1976}).

\bibitem[{\citenamefont{Berger}(1972)}]{Berger72}
\bibinfo{author}{\bibfnamefont{L.}~\bibnamefont{Berger}},
  \bibinfo{journal}{Phys. Rev. B} \textbf{\bibinfo{volume}{5}},
  \bibinfo{pages}{1862} (\bibinfo{year}{1972}).

\bibitem[{\citenamefont{Hirsch}(1999)}]{Hirsch99}
\bibinfo{author}{\bibfnamefont{J.~E.} \bibnamefont{Hirsch}},
  \bibinfo{journal}{Phys. Rev. Lett.} \textbf{\bibinfo{volume}{83}},
  \bibinfo{pages}{1834} (\bibinfo{year}{1999}).

\bibitem[{\citenamefont{Silsbee et~al.}(1979)\citenamefont{Silsbee, Janossy,
  and Monod}}]{Silsbee79}
\bibinfo{author}{\bibfnamefont{R.~H.} \bibnamefont{Silsbee}},
  \bibinfo{author}{\bibfnamefont{A.}~\bibnamefont{Janossy}}, \bibnamefont{and}
  \bibinfo{author}{\bibfnamefont{P.}~\bibnamefont{Monod}},
  \bibinfo{journal}{Phys. Rev. B} \textbf{\bibinfo{volume}{19}},
  \bibinfo{pages}{4382} (\bibinfo{year}{1979}).

\bibitem[{\citenamefont{Tserkovnyak et~al.}(2002)\citenamefont{Tserkovnyak,
  Brataas, and Bauer}}]{Tserkovnyak02}
\bibinfo{author}{\bibfnamefont{Y.}~\bibnamefont{Tserkovnyak}},
  \bibinfo{author}{\bibfnamefont{A.}~\bibnamefont{Brataas}}, \bibnamefont{and}
  \bibinfo{author}{\bibfnamefont{G.~E.~W.} \bibnamefont{Bauer}},
  \bibinfo{journal}{Phys. Rev. Lett.} \textbf{\bibinfo{volume}{88}},
  \bibinfo{eid}{117601} (\bibinfo{year}{2002}).

\bibitem[{\citenamefont{Saitoh et~al.}(2006)\citenamefont{Saitoh, Ueda,
  Miyajima, and Tatara}}]{Saitoh06}
\bibinfo{author}{\bibfnamefont{E.}~\bibnamefont{Saitoh}},
  \bibinfo{author}{\bibfnamefont{M.}~\bibnamefont{Ueda}},
  \bibinfo{author}{\bibfnamefont{H.}~\bibnamefont{Miyajima}}, \bibnamefont{and}
  \bibinfo{author}{\bibfnamefont{G.}~\bibnamefont{Tatara}},
  \bibinfo{journal}{Appl. Phys. Lett.} \textbf{\bibinfo{volume}{88}},
  \bibinfo{pages}{182509} (\bibinfo{year}{2006}).

\bibitem[{\citenamefont{Uchida et~al.}(2008)\citenamefont{Uchida, Takahashi,
  Harii, Ieda, Koshibae, Ando, Maekawa, and Saitoh}}]{Uchida08}
\bibinfo{author}{\bibfnamefont{K.}~\bibnamefont{Uchida}},
  \bibinfo{author}{\bibfnamefont{S.}~\bibnamefont{Takahashi}},
  \bibinfo{author}{\bibfnamefont{K.}~\bibnamefont{Harii}},
  \bibinfo{author}{\bibfnamefont{J.}~\bibnamefont{Ieda}},
  \bibinfo{author}{\bibfnamefont{W.}~\bibnamefont{Koshibae}},
  \bibinfo{author}{\bibfnamefont{K.}~\bibnamefont{Ando}},
  \bibinfo{author}{\bibfnamefont{S.}~\bibnamefont{Maekawa}}, \bibnamefont{and}
  \bibinfo{author}{\bibfnamefont{E.}~\bibnamefont{Saitoh}},
  \bibinfo{journal}{Nature} \textbf{\bibinfo{volume}{455}}, \bibinfo{pages}{778
  } (\bibinfo{year}{2008}).

\bibitem[{\citenamefont{Berger}(1979)}]{Berger79}
\bibinfo{author}{\bibfnamefont{L.}~\bibnamefont{Berger}}, \bibinfo{journal}{J.
  Appl. Phys.} \textbf{\bibinfo{volume}{50}}, \bibinfo{pages}{7698}
  (\bibinfo{year}{1979}).

\bibitem[{\citenamefont{Berger}(1985)}]{Berger85}
\bibinfo{author}{\bibfnamefont{L.}~\bibnamefont{Berger}},
  \bibinfo{journal}{Journal of Applied Physics} \textbf{\bibinfo{volume}{58}},
  \bibinfo{pages}{450} (\bibinfo{year}{1985}).


\bibitem[{\citenamefont{Jen and Berger}(1986)}]{Jen_a86}
\bibinfo{author}{\bibfnamefont{S.~U.} \bibnamefont{Jen}} \bibnamefont{and}
  \bibinfo{author}{\bibfnamefont{L.}~\bibnamefont{Berger}},
  \bibinfo{journal}{Journal of Applied Physics} \textbf{\bibinfo{volume}{59}},
  \bibinfo{pages}{1278} (\bibinfo{year}{1986}).

\bibitem[{\citenamefont{Hatami et~al.}(2007)\citenamefont{Hatami, Bauer, Zhang,
  and Kelly}}]{Hatami07}
\bibinfo{author}{\bibfnamefont{M.}~\bibnamefont{Hatami}},
  \bibinfo{author}{\bibfnamefont{G.~E.~W.} \bibnamefont{Bauer}},
  \bibinfo{author}{\bibfnamefont{Q.}~\bibnamefont{Zhang}}, \bibnamefont{and}
  \bibinfo{author}{\bibfnamefont{P.~J.} \bibnamefont{Kelly}},
  \bibinfo{journal}{Phys. Rev. Lett.} \textbf{\bibinfo{volume}{99}},
  \bibinfo{pages}{066603} (\bibinfo{year}{2007}).

\bibitem[{\citenamefont{Kovalev et~al.}(2009)\citenamefont{Kovalev,
  Tserkovnyak, V\'{y}born\'{y}, and Sinova}}]{Kovalev09}
\bibinfo{author}{\bibfnamefont{A.~A.} \bibnamefont{Kovalev}},
  \bibinfo{author}{\bibfnamefont{Y.}~\bibnamefont{Tserkovnyak}},
  \bibinfo{author}{\bibfnamefont{K.}~\bibnamefont{V\'{y}born\'{y}}},
  \bibnamefont{and} \bibinfo{author}{\bibfnamefont{J.}~\bibnamefont{Sinova}},
  \bibinfo{journal}{Phys. Rev. B} \textbf{\bibinfo{volume}{79}},
  \bibinfo{eid}{195129} (\bibinfo{year}{2009}).

\bibitem[{\citenamefont{Bauer et~al.}(2010)\citenamefont{Bauer, Bretzel,
  Brataas, and Tserkovnyak}}]{Bauer10}
\bibinfo{author}{\bibfnamefont{G.~E.~W.} \bibnamefont{Bauer}},
  \bibinfo{author}{\bibfnamefont{S.}~\bibnamefont{Bretzel}},
  \bibinfo{author}{\bibfnamefont{A.}~\bibnamefont{Brataas}}, \bibnamefont{and}
  \bibinfo{author}{\bibfnamefont{Y.}~\bibnamefont{Tserkovnyak}},
  \bibinfo{journal}{Phys. Rev. B} \textbf{\bibinfo{volume}{81}},
  \bibinfo{pages}{024427} (\bibinfo{year}{2010}).

\bibitem[{\citenamefont{van Son et~al.}(1987)\citenamefont{van Son, van Kempen,
  and Wyder}}]{Son87}
\bibinfo{author}{\bibfnamefont{P.~C.} \bibnamefont{van Son}},
  \bibinfo{author}{\bibfnamefont{H.}~\bibnamefont{van Kempen}},
  \bibnamefont{and} \bibinfo{author}{\bibfnamefont{P.}~\bibnamefont{Wyder}},
  \bibinfo{journal}{Phys. Rev. Lett.} \textbf{\bibinfo{volume}{58}},
  \bibinfo{pages}{2271} (\bibinfo{year}{1987}).

\bibitem[{\citenamefont{Valet and Fert}(1993)}]{Valet93}
\bibinfo{author}{\bibfnamefont{T.}~\bibnamefont{Valet}} \bibnamefont{and}
  \bibinfo{author}{\bibfnamefont{A.}~\bibnamefont{Fert}},
  \bibinfo{journal}{Phys. Rev. B} \textbf{\bibinfo{volume}{48}},
  \bibinfo{pages}{7099} (\bibinfo{year}{1993}).

\bibitem[{\citenamefont{Tatara et~al.}(2008)\citenamefont{Tatara, Kohno, and
  Shibata}}]{TKS_PR08}
\bibinfo{author}{\bibfnamefont{G.}~\bibnamefont{Tatara}},
  \bibinfo{author}{\bibfnamefont{H.}~\bibnamefont{Kohno}}, \bibnamefont{and}
  \bibinfo{author}{\bibfnamefont{J.}~\bibnamefont{Shibata}},
  \bibinfo{journal}{Physics Reports} \textbf{\bibinfo{volume}{468}},
  \bibinfo{pages}{213} (\bibinfo{year}{2008}).

\bibitem[{\citenamefont{Takeuchi et~al.}(2010)\citenamefont{Takeuchi, Hosono,
  and Tatara}}]{Takeuchi10}
\bibinfo{author}{\bibfnamefont{A.}~\bibnamefont{Takeuchi}},
  \bibinfo{author}{\bibfnamefont{K.}~\bibnamefont{Hosono}}, \bibnamefont{and}
  \bibinfo{author}{\bibfnamefont{G.}~\bibnamefont{Tatara}},
  \bibinfo{journal}{Phys. Rev. B} \textbf{\bibinfo{volume}{81}},
  \bibinfo{pages}{144405} (\bibinfo{year}{2010}).

\bibitem[{\citenamefont{Nakabayashi et~al.}(2010)\citenamefont{Nakabayashi,
  Takeuchi, Hosono, Taguchi, and Tatara}}]{Nakabayashi10}
\bibinfo{author}{\bibfnamefont{N.}~\bibnamefont{Nakabayashi}},
  \bibinfo{author}{\bibfnamefont{A.}~\bibnamefont{Takeuchi}},
  \bibinfo{author}{\bibfnamefont{K.}~\bibnamefont{Hosono}},
  \bibinfo{author}{\bibfnamefont{K.}~\bibnamefont{Taguchi}}, \bibnamefont{and}
  \bibinfo{author}{\bibfnamefont{G.}~\bibnamefont{Tatara}}
  (\bibinfo{year}{2010}).

\bibitem[{\citenamefont{Uchida et~al.}(2009)\citenamefont{Uchida, Takahashi,
  Ieda, Harii, Ikeda, Koshibae, Maekawa, and Saitoh}}]{Uchida09}
\bibinfo{author}{\bibfnamefont{K.}~\bibnamefont{Uchida}},
  \bibinfo{author}{\bibfnamefont{S.}~\bibnamefont{Takahashi}},
  \bibinfo{author}{\bibfnamefont{J.}~\bibnamefont{Ieda}},
  \bibinfo{author}{\bibfnamefont{K.}~\bibnamefont{Harii}},
  \bibinfo{author}{\bibfnamefont{K.}~\bibnamefont{Ikeda}},
  \bibinfo{author}{\bibfnamefont{W.}~\bibnamefont{Koshibae}},
  \bibinfo{author}{\bibfnamefont{S.}~\bibnamefont{Maekawa}}, \bibnamefont{and}
  \bibinfo{author}{\bibfnamefont{E.}~\bibnamefont{Saitoh}},
  \bibinfo{journal}{J. Appl. Phys.} \textbf{\bibinfo{volume}{105}},
  \bibinfo{eid}{07C908} (\bibinfo{year}{2009}).

\end{thebibliography}
\end{document}